\documentclass[sigconf, prologue, x11names, rgb]{acmart}
\pdfoutput=1

\usepackage{booktabs} 
\usepackage[utf8]{inputenc}

\usepackage{solarized}
\usepackage{tikz}

\usepackage{xspace}
\usepackage{lstautogobble}  
\usepackage{zi4}            

\usepackage{float}
\usepackage{listings}

\usepackage{multirow}

\usepackage{comment}
\includecomment{fullversion}\excludecomment{confversion}

\newfloat{lstfloat}{tbp}{lop}
\floatname{lstfloat}{Listing}

\setcopyright{acmlicensed}

\newcommand{\code}[1]{\texttt{#1}}

\newcommand{\confversioncmd}[1]{}
\begin{confversion}
  \renewcommand{\confversioncmd}[1]{#1}
\end{confversion}
\newcommand{\fullversioncmd}[1]{}
\begin{fullversion}
  \renewcommand{\fullversioncmd}[1]{#1}
\end{fullversion}

\newcommand{\name}{\textsc{zksk}\xspace}

\newcommand{\paranodot}[1]{\vspace{1mm}\noindent\textbf{#1}}
\newcommand{\para}[1]{\vspace{1mm}\noindent\textbf{#1}.}

\renewcommand{\G}{\mathbb{G}}
\newcommand{\generator}{g}
\newcommand{\grouporder}{q}
\newcommand{\Zq}{\mathbb{Z}_q}
\newcommand{\randin}{\in_R}

\newcommand{\randomizer}[1]{r_{#1}}
\newcommand{\commit}[1]{R_{#1}}
\newcommand{\challenge}{c}
\newcommand{\response}[1]{s_{#1}}

\newcommand{\tikzmagicarrow}[3]{
\begin{tikzpicture}[]
  \draw[#1](0,0) -- node[above=-0.5ex]{\ensuremath{#2}} (#3,0);
  \node[draw=none] (bottom) at (0,-0.5ex) {};
  \node[draw=none] (top) at (0,1ex) {};
  \node[draw=none] (lowerleft) at (bottom-|current bounding box.west) {};
  \node[draw=none] (topright) at (top-|current bounding box.east) {};
  \pgfresetboundingbox
  \draw[draw=none,use as bounding box] (lowerleft) rectangle (topright);
\end{tikzpicture}
}
\newcommand{\tikzlongarrow}[2]{\tikzmagicarrow{#1}{#2}{1.5}}

\newcommand{\diagramsend}[1]{\tikzlongarrow{->}{#1}}
\newcommand{\diagramrecv}[1]{\tikzlongarrow{<-}{#1}}

\renewcommand\footnotetextcopyrightpermission[1]{} 
\begin{document}
\fancyhead{}

\title{\name: A Library for Composable Zero-Knowledge Proofs}

\author{Wouter Lueks}
\affiliation{
  \institution{EPFL SPRING Lab}
}
\email{wouter.lueks@epfl.ch}

\author{Bogdan Kulynych}
\affiliation{
  \institution{EPFL SPRING Lab}
}
\email{bogdan.kulynych@epfl.ch}

\author{Jules Fasquelle}
\affiliation{
  \institution{EPFL}
}
\email{jules.fasquelle@epfl.ch}

\author{Simon Le Bail-Collet}
\affiliation{
  \institution{EPFL}
}
\email{simon.lebail-collet@epfl.ch}

\author{Carmela Troncoso}
\affiliation{
  \institution{EPFL SPRING Lab}
}
\email{carmela.troncoso@epfl.ch}

\begin{abstract}
Zero-knowledge proofs are an essential building block in many privacy-preserving
systems. However, implementing these proofs is tedious and error-prone.
In this paper, we present \name, a well-documented Python library for defining and
computing sigma protocols: the most popular class of zero-knowledge proofs.
In \name, proofs compose: programmers can convert smaller proofs into
building blocks that then can be combined into bigger
proofs.
\name features a modern Python-based domain-specific language. This makes
possible to define proofs without learning a new custom language, and to benefit
from the rich Python syntax and ecosystem.
\end{abstract}

\copyrightyear{2019}
\acmYear{2019}
\acmConference[WPES'19]{18th Workshop on Privacy in the Electronic Society}{November 11, 2019}{London, United Kingdom}
\acmBooktitle{18th Workshop on Privacy in the Electronic Society (WPES'19), November 11, 2019, London, United Kingdom}
\acmPrice{15.00}
\acmDOI{10.1145/3338498.3358653}
\acmISBN{978-1-4503-6830-8/19/11}

\maketitle

\section{Introduction}
Privacy-preserving systems use zero-knowledge proofs to prove that outputs have been computed correctly,
without revealing sensitive information about inputs. In online voting systems, voters prove that they correctly encrypted
their vote, without revealing any information about the selected candidate~\cite{AdidaMPQ09}. In
anonymous authentication systems, users prove that they have access to a resource, without revealing
any information that could reveal their identity or make accesses linkable~\cite{DavidsonGSTV18,HenryG13}.

Implementing zero-knowledge proofs is tedious. Academic papers often use high-level
Camenisch-Stadler notation~\cite{CamenischS97} to succinctly specify the intent of the proofs.
The concrete implementations of proofs, however, often require hundreds or thousands lines of code. 
While not necessarily difficult, these implementations are tedious 
and require implementing the same primitives repeatedly.

Implementations are also error-prone. To simplify deployment, most systems use the Fiat-Shamir
heuristic~\cite{FiatS86}. Incorrectly applying this heuristic, however, can lead to serious
vulnerabilities. For instance, in both the Helios and SwissPost/Scytl voting systems, incorrect
application of the Fiat-Shamir heuristic led to accepting incorrect encrypted
votes~\cite{BernhardPW12,LewisPT19outcome,LewisPT19outcomeaddendum}. 

We propose \name, the \emph{Zero-Knowledge Swiss Knife}, a Python library for defining and
computing sigma protocols -- the most popular class of zero-knowledge proofs. The library provides a
simple API to define proofs inspired by the Camenisch-Stadler notation~\cite{CamenischS97}.
Additionally, the \name library protects the programmer against mistakes.
It applies the Fiat-Shamir construction correctly
automatically; and it refuses to compute or-proofs that would reveal secrets.
Consider the zero-knowledge proof that an additive ElGamal
ciphertext~\cite{ElGamal85}
$(c_1, c_2) = (g^r, g^m \cdot h^r)$ for a public key $h$ encrypts the value $m =
0$ or $m = 1$. In Camenisch-Stadler notation, we would write:
\begin{equation*}
  PK\{ (r) : (c_1 = g^r \land c_2 = h^r) \lor (c_1 = g^r \land c_2 g^{-1} = h^r) \}
\end{equation*}
to denote the proof of knowledge of a secret $r$ such that the
expression after the colon holds.
In our library we write (using additive notation, and
\code{G} and \code{H} for $g$ and $h$):
\begin{lstlisting}
r = Secret()
enc0 = DLRep(c1, r * G) & DLRep(c2, r * H)
enc1 = DLRep(c1, r * G) & DLRep(c2 - G, r * H)
stmt = enc0 | enc1
\end{lstlisting}
to define the statement \code{stmt} of the same proof.

Systems often compose simpler zero-knowledge proofs to achieve their purpose. In voting systems,
the voter's vote is usually represented by a vector of ciphertexts, each corresponding to a
candidate. To ensure correctness, voters prove that each ciphertext encrypts a bit \emph{and} that
not too many bits are set. In anonymous authentication systems, users prove that they have a
credential \emph{and} that this credential has not yet been revoked.
%
Zero-knowledge proofs defined in \name support composition. For example, the statement $\code{stmt}$ in the
example composes the two disjuncts $\code{enc0}$ and $\code{enc1}$.

The \name library also makes it easy to define new
building blocks that can themselves be composed. For example, \name already
defines a range-proof construction. Consider a voting scheme in which the voter
can select at most 5 candidates. Then the voter must show that the sum of
votes, another ElGamal ciphertext $(c_1, c_2)$, encrypts $m$ such that $0 \leq m < 5$.
We express this in \name as:
\begin{lstlisting}
r = Secret()
m = Secret()
enc_stmt = DLRep(c1, r * G) & DLRep(c2, m * G + r * H)
# Prove that c2 commits to m (bases G, H) and 0 <= m < 5
range_stmt = RangeStmt(c2, G, H, 0, 5, m, r)
stmt = enc_stmt & range_stmt
\end{lstlisting}

\para{Existing libraries and compilers}
In this work, we focus on sigma protocols. The tools for defining and
computing generic zero-knowledge proofs, such as
zk-SNARKs~\cite{Ben-SassonCTV14}, are thus out of scope.
See Table~\ref{tab:comparison} for a comparison of sigma-protocol libraries.

The \emph{Secure Computation API} (SCAPI)~\cite{EjgenbergFLL12,SCAPI} is a C++
library that provides a small set of sigma-protocol primitives. Programmers write C++ code to define and compute
proofs. The high-level interface is well-documented, but
to use individual primitives programmers must read the source code.
Primitives can be composed using AND and OR
constructions, but SCAPI's composition is limited: the
programmer cannot specify that the \emph{same} variable occurs in multiple
conjuncts.
Instead, the programmer must define a new primitive from scratch.
The \emph{emmy} Go library also implements several sigma-protocol 
primitives. Programmers write Go code to define and compute proofs, but the
primitives cannot be combined with conjunctions or disjunctions to form bigger
proofs, and documentation is minimal.



\emph{YAZKC} (Yet Another Zero-Knowledge Compiler)~\cite{AlmeidaBBKSS10} and
\emph{Cashlib}~\cite{MeiklejohnEKHL10} instead use a custom language for defining
zero-knowledge proofs, and provide a compiler that transforms the specification
into code to compute and verify proofs. The YAZKC and Cashlib DSLs
resembles the notation of Camenisch and Stadler. Cashlib does not support OR
constructions, YAZKC does.  In both cases, the DSL does not support defining new
high-level building blocks, requiring proofs
instead to be (re)written in their entirety in terms of the DSL's building blocks.
We could not fully evaluate YAZKC as the source is no longer available online.
Both YAZKC and Cashlib support hidden order groups. Our \name library does not,
as these are nowadays usually replaced by pairings, which we do support.
%
The \emph{zkp} Rust library~\cite{dalek} provides a Rust DSL to
define simple proofs. These proofs cannot be combined in disjunctions or conjunctions.

\begin{table}[tbp]
  \newcommand{\yes}{\checkmark}
  \newcommand{\no}{$\times$}
  \caption{\label{tab:comparison}Comparison between different zero-knowledge
    proof libraries. Columns:
    \emph{AND, OR} -- support for conjunct and disjunct statements,
    \emph{Composing (Cmp)} -- defined statements compose into bigger statements,
    \emph{Interactive (Int)} -- interactive prove/verification mode,
    \emph{FS} -- non-interactive proofs through Fiat-Shamir heuristic,
    \emph{Language (Lang)} -- language in which the tool is implemented,
    \emph{DSL} -- language in which proofs can be defined,
    \emph{Documentation (Docs)} -- available
    documentation.}
  {\small
  \begin{tabular}{@{}l@{}cccccccc@{}}
    \toprule
    & AND & OR & Cmp & Int & FS & Lang & DSL &  Docs \\
    \midrule
    SCAPI~\cite{EjgenbergFLL12,SCAPI} & \yes & \yes & \textasciitilde & \yes & \yes & C++ & C++ & \yes \\
    Emmy~\cite{emmy} & \no & \no & \no & \yes & \no & Go & - & min. \\
    YAZKC~\cite{AlmeidaBBKSS10} & \yes & \yes & \no & \yes & \yes & C & Custom & \textasciitilde \\
    Cashlib~\cite{MeiklejohnEKHL10} & \yes & \no & \no & \no & \yes & C++ & Custom & \textasciitilde \\
    zkp~\cite{dalek} & \yes & \no & \no & \no & \yes & Rust & Rust & min. \\
    \bf{\name} &  \yes & \yes & \yes & \yes & \yes & Python & Python & \yes \\
    \bottomrule
  \end{tabular}
  }
\end{table}

\para{Contributions} In this paper we make the following contributions.

\paranodot{\checkmark} We present \name, a well-documented Python library that provides an
API for defining and computing zero-knowledge proofs. 

\paranodot{\checkmark} \name protects programmers against common errors and supports full composition (conjunctions
and disjunctions). It comes with useful building
blocks that can be used to instantiate many existing zero-knowledge proofs: 
proofs of signatures, range proofs, and inequality proofs.
Users can also define their own building blocks. 

\section{Background}
Throughout this paper, let $\G$ be a cyclic group of prime order $\grouporder$
generated by $\generator$. Let $\Zq$ be the integers modulo $\grouporder$. We
write $a \parallel b$ for the concatenation of two strings, and $a \randin A$
to denote that $a$ is drawn uniformly at random from the finite set $A$. We call
an expression $C = g_1^{x_1} \cdots g_n^{x_n}$ a \emph{discrete logarithm
representation} of $C$ with respect to the bases $g_1, \ldots, g_n$.

\begin{figure}[h!]
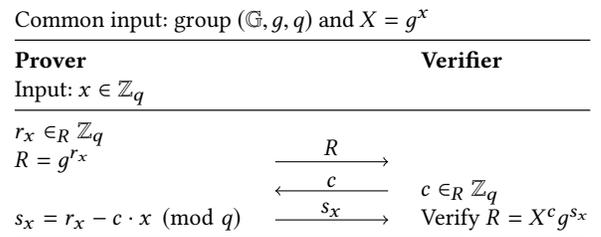

  \begin{tabular}{@{}lcl@{}}
    \toprule
    \multicolumn{3}{@{}l}{Common input: group $(\G, \generator, \grouporder)$ and $X = g^x$} \\
    \midrule
    \textbf{Prover} & & \textbf{Verifier} \\
    Input: $x \in \Zq$ \\
    \midrule
    $\randomizer{x} \randin \Zq$ \\
    $\commit{} = g^{\randomizer{x}}$ & \diagramsend{\commit{}} & \\
     & \diagramrecv{\challenge} &  $\challenge \randin \Zq$ \\
    $\response{x} = \randomizer{x} - \challenge \cdot x \pmod{\grouporder}$ & \diagramsend{\response{x}} & Verify $\commit{} = X^{\challenge} g^{\response{x}}$ \\
    \bottomrule
  \end{tabular}
  \vspace{-3mm}
  \caption{\label{fig:schnorr}Schnorr's proof of identity. A simple sigma
    protocol that proves knowledge of $x$ such that $X = g^x$.}
\end{figure}

In this paper, we focus on sigma protocols~\cite{Damgard10}: 3-move zero-knowledge
proofs~\cite{GoldwasserMR89}, that provide \emph{honest-verifier zero-knowledge}.
The most well-known example is Schnorr's proof of
identification~\cite{Schnorr89}, see Figure~\ref{fig:schnorr}. It proceeds in
three phases: (1) the prover sends a \emph{commitment}, the value $\commit{}$; (2) the verifier
sends a \emph{challenge} $\challenge$; and (3) the prover sends a \emph{response} $\response{x}$.
Finally, the verifier checks the correctness of the response. Every sigma protocol follows this
structure.
In Camenisch-Stadler notation~\cite{CamenischS97}, we express the proof
statement as $PK\{ (x) : X = g^x \}$ to denote that the prover proves knowledge
of the (secret) value $x$ such that $X = g^x$.

The protocol is zero-knowledge because transcripts $(\commit{}, \challenge, \response{x})$ that are
accepted by the verifier can be \emph{simulated} without knowledge of the secret~\cite{Schnorr89}.
Intuitively, the verifier can therefore not convince anybody else of the veracity of the statement.


Sigma protocols can be combined using conjunctions, e.g., prove knowledge of the
discrete logarithms of $X_1$ \emph{and} $X_2$ (with respect to $g_1$ and $g_2$
), and in fact that they are the same:
\begin{equation*}
  PK\{ (x) : X_1 = g_1^x \land X_2 = g_2^x \}.
\end{equation*}
To do so, run two Schnorr identification protocols in parallel, using the \emph{same} randomizer for
the secret $x$. This approach enables proving arbitrary conjunctions.
\confversioncmd{See the full paper~\cite{fullversion} for the details.}
\fullversioncmd{See Figure~\ref{fig:equal} in the appendix for the details.}

Sigma protocols can also be combined using disjunctions, e.g., we can prove knowledge of the
discrete logarithm of $X_1$ \emph{or} $X_2$:
\begin{equation*}
  PK\{ (x) : X_1 = g_1^x \lor X_2 = g_2^x \}.
\end{equation*}
The OR construction~\cite{Damgard10}, simulates the untrue disjunct, while honestly proving the true disjunct.
\confversioncmd{See the full paper~\cite{fullversion} for the details.}
\fullversioncmd{We provide the detailed protocol in Figure~\ref{fig:or} in appendix.}

The Fiat-Shamir heuristic~\cite{FiatS86} turns interactive protocols into non-interactive
proofs. With this heuristic, the prover computes the challenge $\challenge$ by hashing its
commitments together with the proof statement. For Schnorr's protocol in
Figure~\ref{fig:schnorr}, she would compute $\challenge = H(R \parallel X)$. Including the proof
statement (in this case including $X$ suffices) is essential, lest the prover can fake proofs~\cite{BernhardPW12}.

\section{\name Design And Implementation}

In this section we overview the core functionalities of \name from a user's
perspective, and then outline how they are implemented.
Code examples are written in Python. The library is open source and is extensively
documented.\footnote{\url{https://github.com/spring-epfl/zksk}} The compiler relies on the \code{petlib}\footnote{\url{https://github.com/gdanezis/petlib}} Python
bindings to OpenSSL to support elliptic curves and pairings.

\subsection{Components}

\para{Discrete logarithm representations}
The \name library makes it easy to express equations about discrete
logarithm representations, the basic building block of sigma protocols.
Listing~\ref{lst:simple_example} shows how to express the 
statement $PK\{ (x,r) : C = g^x h^r \}$, and how to construct and verify the
corresponding proof. We assume the values \code{C}, \code{G}, and \code{H} are
defined and in scope.
First, the prover defines the values of which it will
prove knowledge (lines 2--3). Note that it passes in the real values of the secrets.
Then it expresses the proof statement (line 4). The first argument of
\code{DLRep} is the left-hand side of the discrete logarithm representation, the
right-hand side expresses the left-hand side in terms of the bases and secrets.
Finally, it constructs the (non-interactive) proof (line 5).
The verifier first defines the proof statement (lines 8--9), and uses it to verify the proof (line 10). 

\begin{lstfloat}[h!]
\begin{lstlisting}[numbers=left, xleftmargin=2.4em]
  # Prover
  x = Secret(20)
  r = Secret(1337)
  stmt = DLRep(C, x * G + r * H)
  proof = stmt.prove()

  # Verifier
  x_prime, r_prime = Secret(), Secret()
  stmt_prime = DLRep(C, x_prime * G + r_prime * H)
  assert stmt_prime.verify(proof)
\end{lstlisting}
\vspace{-4mm}
\caption{\label{lst:simple_example} Using \name to prove and verify the
  statement $PK\{ (x,r) : C = g^xh^r \}$. In the code \code{C} is the commitment
  $C$.}
\end{lstfloat}

\para{Conjunctions and disjunctions}
We can combine statements into conjunctions or disjunctions.
The \code{\&} (and) operator combines statements into a conjunction:
the library will prove that secrets that appear in
multiple statements are the same.
Similarly, the \code{|} (or) operator combines statements into a
disjunction. Checking which disjunct is true is computationally
expensive. Thus, \name requires the prover to indicate
whether a disjunct is true or simulated:
\begin{lstlisting}
  x  = Secret(12345)
  stat = DLRep(X1, x * G1, simulated=True) |
         DLRep(X1, x * G2, simulated=False)
\end{lstlisting}




\para{Defining and using primitives}
The \name library includes several useful primitives to define more complicated
statements: proofs of knowledge of a BBS+ signature~\cite{AuSM06}, inequality of
discrete logarithms~\cite{HenryG13}, and range
proofs~\cite{BellareG97,Schoenmakers05}. The syntax is as follows:
  \begin{lstlisting}
  # Possession of BBS+ signature over messages:
  msgs = [Secret() for _ in range(4)]
  stmt1 = BBSPlusSignatureProof(msgs, pk)

  # Inequality of discrete logs (see below):
  stmt2 = DLRepNotEqual([Y1, G1], [Y2, G2])

  # Let com = x * G + r * H be a commitment to x
  # Proof that x lies in the range [a, b):
  x, r = Secret(), Secret()
  stmt3 = RangeStmt(com, G, H, a, b, x, r)
  \end{lstlisting}

Users can easily define new primitives.
These primitives could require extra computations and verifications.
We take as an example the \emph{proof of
inequality of two discrete logarithms} by Henry and Goldberg~\cite{HenryG13},
proving the statement $PK\{ (x) : H_0 = h_0^x \land H_1 \neq h_1^x \}$. 
This statement cannot be directly translated into the primitives we have defined
before. Instead, we follow Henry and Goldberg's approach. The prover first picks a
randomizer $r \randin \Zq$ and computes the value $C = (h_1^x / H_1)^r$, and then
proves:
\begin{equation*}
  PK\left\{ (\alpha, \beta) : 1 = h_0^\alpha H_0^\beta \land C = h_1^\alpha H_1^\beta \right\}
\end{equation*}
where $\alpha = xr \mod{\grouporder}$ and $\beta = -r \mod{\grouporder}$.
When verifying the proof, the verifier needs to additionally check that $C \neq 1$.

Primitives in \name extend the \code{ExtendedProofStmt} class, provide a
constructor, and override the \code{construct\_stmt} to return a proof
statement. Moreover, they can override \code{precommit} to compute a
pre\-commitment, and \code{validate} to perform post-val\-idation.

Listing~\ref{lst:dlnotequal} shows how to implement the \code{DLNotEqual} proof. 
First, we define the constructor. It stores the arguments (lines
3--5), computes some convenience values (lines 7--8), and defines the secrets
\code{alpha} and \code{beta} (line 11). Then, we override the \code{precommit}
function to compute the value $C$ that acts as a precommitment (lines 13--21).
This function also sets the values of the secrets \code{alpha} and \code{beta}.
The function \code{construct\_proof} returns the proof statement defined
above (lines 23--29). Finally, \code{validate} (lines 31--33) verifies that
$C$ is not the unity element.

\begin{lstfloat}[h!]
\begin{lstlisting}[numbers=left, xleftmargin=2.5em]
class DLNotEqual(ExtendedProofStmt):
  def __init__(self, valid_pair, invalid_pair, x):
    self.lhs = [valid_pair[0], invalid_pair[0]]
    self.bases = [valid_pair[1], invalid_pair[1]]
    self.x = x

    self.infty = self.bases[0].group.infinite()
    self.order = self.bases[0].group.order()

    # The internal proof uses two constructed secrets
    self.alpha, self.beta = Secret(), Secret()

  def precommit(self):
    blinder = self.order.random()

    # Set the value of the two internal secrets
    self.alpha.value = self.x.value * blinder % order
    self.beta.value = -blinder % order

    precommitment = blinder * (self.x.value * self.bases[1] - self.lhs[1])
    return precommitment

  def construct_stmt(self, precom):
    p1 = DLRep(infty,
               self.alpha * self.bases[0] +
               self.beta * self.lhs[0])
    p2 = DLRep(precom, self.alpha * self.bases[1] +
                       self.beta * self.lhs[1])
    return p1 & p2

  def validate(self, precommitment):
    if self.precommitment == self.infty:
      raise ValidationError("Invalid precommitment")
\end{lstlisting}
\vspace{-4mm}
\caption{\label{lst:dlnotequal}Full implementation of the \code{DLNotEqual} primitive.}
\end{lstfloat}

New primitives created by extending \code{ExtendedProofStmt} compose as any other proof
statement. However, they cannot be themselves used in the constructed
proof of other new primitives using
\code{ExtendedProofStmt}. We aim to add this functionality soon.

\begin{lstfloat}
{\small
\begin{tabular}{@{}rl@{}}
  \toprule
  & \textbf{Input:} a (compositional) proof statement \code{stmt}  \\
  & \textbf{Output:} a non-interactive $\pi = (\code{precommitment}, \code{chal}, \code{resp})$  \\
  \midrule
  1 & Recursively call \code{precommit()} on all parts of \code{stmt} \\
  2 & Let \code{precommitment} be the combined precommitments \\
  3 & Create constructed proofs for all parts of \code{stmt} \\
  4 & Verify that secrets inside OR clause do not appear elsewhere \\
  5 & Let ${\code{sec}_1, \ldots, \code{sec}_n}$ be the unique secrets in all
  parts of \code{stmt} \\
  6 & Pick randomizers ${\code{rand}_1, \ldots, \code{rand}_n}$ for the secrets \\
  7 & Compute $\code{commitment}$ for \code{stmt} recursively using $\{\code{rand}_i\}$ \\
  8 & Let $\code{chal} = H(\code{repr(stmt)} \parallel \code{precommitment} \parallel \code{commitment})$ \\
  9 & Compute $\{ \code{resp}_i \}$ for each secret using $\code{chal}$ and $\{\code{rand}_i\}$. \\
  10 & Return $\pi = (\code{precommitment}, \code{chal}, \code{resp}) $ \\
  \bottomrule
\end{tabular}
}
  \vspace{-4mm}
  \caption{\label{lst:compute-proof}Computing a non-interactive proof}
\end{lstfloat}

\subsection{Implementation}

\para{Robust statement identifiers}
The correct application of the Fiat-Shamir heuristic mandates including a
representation of the statement in the hash function. Computing such a
representation, however, is difficult. Consider the two statements:
\begin{lstlisting}
  x, y = Secret(), Secret()
  p1 = DLRep(A, x * G) & DLRep(B, x * H) & DLRep(C, y * Z)
  p2 = DLRep(A, x * G) & DLRep(B, y * H) & DLRep(C, y * Z)
\end{lstlisting}
The \name library must differentiate between the secret \code{x} appearing
twice, and the secret \code{y} appearing twice. Moreover, this method must be
robust even if the prover's and verifier's statement definition execute on
different machines. Hence, we cannot use Python's built-in object identifiers,
as they can change across executions.

One way for differentiating the secrets would be to assign a canonical unique name to each
secret. Our initial experiments showed, however, that manually assigning names to secrets results in
cumbersome code.  Therefore, \name automatically assigns identifiers to secrets in the \emph{order}
in which they first occur in the statement.
\confversioncmd{See the full paper~\cite{fullversion} for the details.}
\fullversioncmd{See Listing~\ref{lst:compute-id} in the appendix for the details.}

\para{Computing proofs}
A call to \code{stmt.prove()}, see Listing~\ref{lst:simple_example}, computes a
non-interactive proof as in Listing~\ref{lst:compute-proof}. First we compute
the precommitments and the concrete constructed proofs for the
custom primitives (lines 1--3). Lines 5--10 then execute the proof, similarly to
Figure~\ref{fig:schnorr}.

The library tries to actively prevent programmer errors. Line 8 applies the
strong Fiat-Shamir heuristic~\cite{BernhardPW12}: it adds the statement's
representation as input to the hash function.

The library also detects dangerous OR proofs. Consider a
simplification of the statement on page 1 that $(c_1, c_2)$ encrypts a bit:
\begin{equation*}
  PK\{ (r) : c_1 = g^r \land (c_2 = h^r \lor c_2 g^{-1} = h^r) \}
\end{equation*}
A na\"ive application of steps 5--10, which picks one randomizer for $r$ and
uses that randomizer in both conjuncts results in a proof that reveals $r$
itself.
Let $e$ be the challenge received from the verifier.
Suppose the first disjunct is true. To prove the full statement, the na\"ive
approach first simulates the second disjunct, obtaining a transcript with a
challenge $e_2$. The challenge for the first disjunct is then $e_1 = e - e_2 \pmod{q}$.
As a result, the na\"ive approach uses the challenge $e$ for the first conjunct
($c_1 = g^r$) and the challenge $e_1$ for the first disjunct ($c_2 = h^r$).
However, given responses for secret $r$ for two different challenges with the same 
randomizer, an attacker can trivially extract $r$, violating the zero-knowledge property.


The \name library prevents this flaw by requiring that secrets that appear in
OR clauses cannot also appear elsewhere.
It is up to
the programmer to resolve this problem when detected. On page 1, we moved the first conjunct
inside, creating a disjunctive normal form. An alternative is to bind the
offending secret to a Pedersen commitment, and then repeat that commitment
inside the OR clause.

\section{Evaluation}

To determine the overhead of using a Python library when computing proofs,
we compare it to the time of computing the proofs with the underlying
cryptographic library.
We first measure the running time of proving knowledge of a BBS+ signature
using \name. We then compute a lower bound on the cost without \name 
by counting the number of group operations the proof takes
and multiplying it by the cost of group operations in the underlying cryptographic library. This lower bound
does not include hash functions nor modular arithmetic. 
We find that 90\% of the running time for \name is due to the group operations. We 
conclude that the overhead of using a
Python library is small.
\confversioncmd{See the full paper~\cite{fullversion} for more details.}
\fullversioncmd{See Appendix~\ref{app:evaluation} for more details.}

We did a small literature study to determine the usefulness of \name.
We explored papers in the last two editions of relevant academic
conferences: PETS, ACM CCS, WPES, and NDSS, and found 7 papers that use sigma
protocols. All of these protocols can be implemented with \name.
\confversioncmd{See the full paper~\cite{fullversion} for more details.}
\fullversioncmd{See Table~\ref{tab:papers} in the appendix for more details.}

\section{Conclusions}

We presented \name, a Python-based library for defining and computing
zero-knowledge proofs based on sigma protocols. Unlike existing libraries, \name
does not rely on a custom language to define proofs, but on an easy-to-use Python-based DSL.
It provides several high-level primitives, and makes it easy to define new high-level primitives,
all of which can be composed to construct bigger proofs.

A small literature study shows that \name is indeed sufficient to
implement sigma protocols encountered in real research papers. We hope that
\name will be a valuable tool to make defining and evaluating such protocols
easier.

\section*{Acknowledgements}
We thank Ian Goldberg and Nick Hopper for pointing out the problem with na\"ively composing OR proofs.

\bibliographystyle{ACM-Reference-Format}
\bibliography{bibliography}

\appendix
\begin{figure}
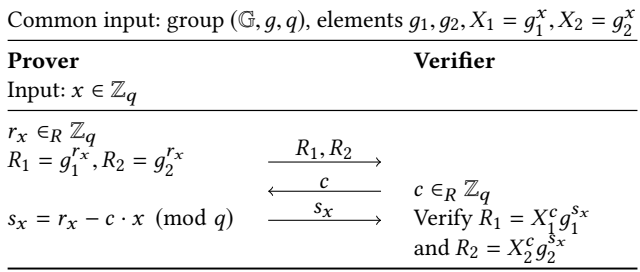

  \begin{tabular}{@{}lcl@{}}
    \toprule
    \multicolumn{3}{@{}l@{}}{Common input: group $(\G, \generator, \grouporder)$, elements $g_1, g_2, X_1 = g_1^x, X_2 = g_2^x$} \\
    \midrule
    \textbf{Prover} & & \textbf{Verifier} \\
    Input: $x \in \Zq$ \\
    \midrule
    $\randomizer{x} \randin \Zq$ \\
    $\commit{1} = g_1^{\randomizer{x}}, \commit{2} = g_2^{\randomizer{x}}$ & \diagramsend{\commit{1},\commit{2}} & \\
     & \diagramrecv{\challenge} &  $\challenge \randin \Zq$ \\
    $\response{x} = \randomizer{x} - \challenge \cdot x \pmod{\grouporder}$ &
      \diagramsend{\response{x}} & Verify $\commit{1} = X_1^{\challenge}
      \generator_1^{\response{x}}$ \\
    & & and $\commit{2} = X_2^{\challenge} \generator_2^{\response{x}}$ \\
    \bottomrule
  \end{tabular}
  \vspace{-3mm}
  \caption{\label{fig:equal} Proof of knowledge of the same discrete logarithm of $X_1$ \emph{and} $X_2$ with respect to $g_1$ and $g_2$ respectively.}
\end{figure}

\begin{figure}
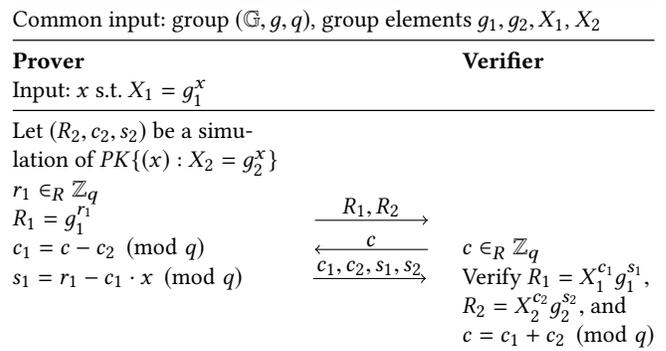

  \begin{tabular}{@{}lcl@{}}
    \toprule
    \multicolumn{3}{@{}l}{Common input: group $(\G, \generator, \grouporder)$, group elements $g_1, g_2, X_1$, $X_2$} \\
    \midrule
    \textbf{Prover} & & \textbf{Verifier} \\
    Input: $x$ s.t. $X_1 = g_1^x$ \\
    \midrule
    Let $(\commit{2}, \challenge_2, \response{2})$ be a simu- \\
    lation of $PK\{ (x) : X_2 = g_2^{x} \}$ \\
    $\randomizer{1} \randin \Zq$ \\
    $\commit{1} = g_1^{\randomizer{1}}$ & \diagramsend{\commit{1},\commit{2}} & \\
    $c_1 = c - c_2 \pmod{\grouporder}$ & \diagramrecv{\challenge} &  $\challenge \randin \Zq$ \\
    $\response{1} = \randomizer{1} - \challenge_1 \cdot x \pmod{\grouporder}$ &
      \diagramsend{\challenge_1, \challenge_2, \response{1}, \response{2}} &
      Verify $\commit{1} = X_1^{\challenge_1} \generator_1^{\response{1}},$ \\
    & & $\commit{2} = X_2^{\challenge_2} \generator_2^{\response{2}},$ and \\
    & & $\challenge = c_1 + c_2 \pmod{\grouporder}$ \\
    \bottomrule
  \end{tabular}
  \vspace{-3mm}
  \caption{\label{fig:or} Proof of knowledge of discrete logairthm of $X_1$
    \emph{or} $X_2$. (Assuming the prover knows $x$ such that $X_1 = g_1^x$)}
\end{figure}

\begin{figure}[!tbp]
  \begin{tabular}{@{}rl@{}}
  \toprule
  & \textbf{Inputs:} \\
  & \quad a compositional proof statement \code{stmt} \\
  & \quad optional mapping of secrets to identifiers \code{secret\_id\_map} \\
  & \textbf{Output:} a robust proof-statement identifier  \\
  \midrule
  1 & If \code{secret\_id\_map} is not given: \\
  2 & \quad Recursively get secret variables from all parts of \code{stmt} \\
  3 & \quad Let \code{secrets} be the combined secret variables\\
  4 & \quad Assign identifiers to \code{secrets} in order of occurrence \\
  5 & \quad let \code{secret\_id\_map} assign these identifiers to secrets. \\
  6 & \quad Recursively call \code{get\_proof\_id} with \code{secret\_id\_map} \\
  7 & \quad \quad on all parts of \code{stmt} \\
  8 & \quad Return the concatenation of obtained proof identifiers \\
  9& Else: \\
  10& \quad Set \code{secret\_ids} according to \code{secret\_id\_map} \\
  11& \quad \quad for each own secret variable. \\
  12& \quad Return identifier (name, bases, \code{secret\_ids})\\
  \bottomrule
\end{tabular}
  \vspace{-4mm}
  \caption{\label{lst:compute-id} \code{get\_proof\_id}: Procedure that computes a robust
    reprsentation of a proof statement, \code{repr(stmt)}.}
\end{figure}

\begin{table*}[!tbp]
  \caption{\label{tab:papers}Overview of papers using sigma protocols in recent
    editions of PETS, ACM CCS, WPES and NDSS. The table lists for each paper, the
    protocol in that paper, which \name primitives it needs to implement it, and
  which composition operations (AND/OR) are needed to define the full proof.}
\begin{tabular}{@{}p{7cm}lllll@{}}
  \toprule
  Title & Year & Conference & Protocol & Primitives & Composition \\
  \midrule
Mesh: A Supply Chain Solution with Locally Private Blockchain Transactions~\cite{AlTawyG19}                    & 2019 & PETS & Membership Proof             & DLRep     & And, Or \\[15pt]
Cryptography for \#MeToo~\cite{KuykendallKR19}                                                                 & 2019 & PETS & ElGamal.Prove                & DLRep     & And     \\
                                                                                                               &      &      & SecShare.Exp                 & DLRep     & And     \\[5pt]
Privacy-Preserving Similar Patient Queries for Combined Biomedical Data~\cite{SalemBHB19}                      & 2019 & PETS & Blinding Correctness         & DLRep     & And     \\[15pt]
Privacy Pass: Bypassing Internet Challenges Anonymously~\cite{DavidsonGSTV18}                                  & 2018 & PETS & Token Signing                & DLRep     & And     \\[15pt]
\multirow{2}{7cm}{Coconut: Threshold Issuance Selective Disclosure Credentials~\cite{SonninoABMD19}} & 2019 & NDSS & IssueCred                    & DLRep     & And     \\
                                                                                                               &      &      & ProveCred                    & DLRep     & And     \\[5pt]
\multirow{2}{7cm}{Solidus: Confidential Distributed Ledger Transactions via PVORM~\cite{CecchettiZJKJS17}}     & 2017 & CCS  & Hidden-Public-Key Signature  & DLRep     & And     \\
                                                                                                               &      &      & 3-Move Range Proof~\cite{BellareG97, Schoenmakers05} & RangeStmt &         \\[5pt]
\multirow{2}{7cm}{ClaimChain: Improving the Security and Privacy of In-band Key Distribution for Messaging~\cite{KulynychLIDT18}}& 2018 & WPES & Claim Correctness            & DLRep     & And     \\
                                                                                                               &      &      & Inter-Block Non-Equivocation & DLRep     & And, Or \\
\bottomrule
\end{tabular}
\end{table*}

\section{Details of evaluation}
\label{app:evaluation}

\para{Overhead in computing proof of knowledge of a BBS+ signature}
We consider a BBS+ signature with 10 messages, and construct a proof of
knowledge of that signature that hides all 10 messages using \name. For a
\name-compatible implementation of cryptographic pairings, we use the
\code{bplib}\footnote{\url{https://github.com/gdanezis/bplib}} library. Proving
takes about 146\,ms, whereas verification takes about 160\,ms.

BBS+ signatures require a pairing setting with groups $G_1, G_2$ and $G_T$.
Based on a manual implementation of the zero-knowledge proof in C, we conclude
that \name must compute 8 exponentiations in $G_1$, 14 exponentiations in $G_T$
and 15 pairing computations to compute a proof; and 6 exponentiations in $G_1$,
15 in $G_T$, and 13 pairings to verify the proof. Based on measurements on the
underlying \code{bplib} library, we estimate the lower bound on the running time
of proving and verifying at 139\,ms and 146\,ms respectively.  Therefore, we
conclude that the raw cryptographic operations account for more than 90\% of the
running time in \name.

\para{Sigma protocols in recent papers} We explored published papers in the last
two years of PETS, ACM CCS, WPES, and NDSS. We found 7 papers that use sigma
protocols. All of them can be implemented using \name.
Privacy Pass~\cite{DavidsonGSTV18}, however, uses an optimized batch
verification protocol that \name currently does not support. The \name library does
support the basic version of the protocol, and can be used to define a new
primitive that supports batch verification.


\end{document}